%% file: gfllr_arxiv.tex
\documentclass[conference]{IEEEtran}
\IEEEoverridecommandlockouts

\usepackage{bm}
\usepackage{lipsum}
\usepackage{enumitem}
\usepackage{numprint}

\usepackage[linesnumberedhidden,lined,ruled]{algorithm2e} 

\usepackage{amsmath,amssymb,amsfonts,accents,mathrsfs, nccmath}
\usepackage{mathtools,array}
\usepackage{cuted}
\usepackage{nicefrac}
\usepackage{scalerel}

\usepackage{multirow}
\usepackage{tabularx}
\usepackage{booktabs}
\usepackage{makecell}
\usepackage{colortbl} 
\usepackage{multicol}
\newcolumntype{C}[1]{>{\centering\arraybackslash}m{#1}} 

\usepackage{float}
\usepackage{graphicx}
\usepackage{textcomp}
\usepackage{rotating} 

\usepackage{pgfplots}
\usepackage{tikz}
\pgfplotsset{compat=newest} 
\newlength\figureheight 
\newlength\figurewidth 
\usetikzlibrary{spy}

\renewcommand{\t}{\left[t\right]}

\def\bphi{{\boldsymbol{\phi}}}

\def\bOmega{{\boldsymbol{\Omega}}}
\def\bv{{\mathbf{v}}}
\def\bV{{\mathbf{V}}}

\def\bD{{\mathbf{D}}}
\def\bX{{\mathbf{X}}}
\def\bx{{\mathbf{x}}}
\def\bH{{\mathbf{H}}}
\def\bh{{\mathbf{h}}}
\def\bY{{\mathbf{Y}}}
\def\by{{\mathbf{y}}}

\def\bR{{\mathbf{R}}}
\def\bw{{\mathbf{w}}}
\def\bW{{\mathbf{W}}}

\def\bphi{{\boldsymbol{\phi}}}
\def\bPhi{{\boldsymbol{\Phi}}}

\definecolor{amethyst}{rgb}{0.6, 0.4, 0.8}
\definecolor{purple}{rgb}{0.49020,0.18039,0.56078}%
\definecolor{mustard}{rgb}{0.92941,0.69020,0.12941}%
\definecolor{gold}{rgb}{0.7607843137,0.6588235294,0}%
\definecolor{orange}{rgb}{0.82353,0.41176,0.11765}%
\definecolor{dark_green}{rgb}{0.00000,0.50196,0.00000}%
\definecolor{lgreen}{rgb}{0.00000,0.7411764706,0.00000}%
\definecolor{wine}{rgb}{0.3725490196,0,0}%
\definecolor{gray}{rgb}{0.50196,0.50196,0.50196}%
\definecolor{mblue}{rgb}{0.00000,0.45098,0.74118}%
\definecolor{lblue}{rgb}{0.5,0.8,0.9}%
\definecolor{lgray}{rgb}{0.65098,0.65098,0.65098}%
\definecolor{llgray}{rgb}{0.85098,0.85098,0.85098}%
\definecolor{lllgray}{rgb}{0.97,0.97,0.97}%
\definecolor{lpink}{rgb}{0.9843137255,0.7333333,0.8862745098}%
\definecolor{mpink}{rgb}{0.933333,0.1490196078,0.6666666667}

\definecolor{modred_VR}{RGB}{192,51,77}
\definecolor{modpink_VR}{RGB}{214, 97, 143}
\definecolor{vsoftorange_VR}{RGB}{243,212,160}
\definecolor{vividorange_VR}{RGB}{241,147,29}
\definecolor{mddarkorange_VR}{RGB}{143,113,91}
\definecolor{modcyan_VR}{RGB}{97, 202, 214}
\definecolor{modlimegreen_VR}{RGB}{97, 214, 110}

\definecolor{vdarkblue_CA}{RGB}{20,50,92}
\definecolor{softblue_CA}{RGB}{83,152,217}
\definecolor{vsoftorange_CA}{RGB}{244,227,177}
\definecolor{vividorange_CA}{RGB}{217,107,12}
\definecolor{darkmodred_CA}{RGB}{165,58,59}
\definecolor{darkmodlimegreen_CA}{RGB}{59,165,58}
\definecolor{darkmodviolet_CA}{RGB}{111,58,165}

\definecolor{Maroon_ST}{RGB}{128,0,0}
\definecolor{Maroon_ST2}{RGB}{180,0,0}
\definecolor{Red_ST}{RGB}{230,25,75}
\definecolor{MedRed_ST}{RGB}{240,65,105}
\definecolor{Pink_ST}{RGB}{250,190,212}
\definecolor{Pink_ST2}{rgb}{0.83922,0.38039,0.56078}
\definecolor{Brown_ST}{RGB}{170,110,40}
\definecolor{Orange_ST}{RGB}{245,130,48}
\definecolor{Olive_ST}{RGB}{128,128,0}
\definecolor{Yellow_ST}{RGB}{255,255,25}
\definecolor{Apricot_ST}{RGB}{255,215,180}
\definecolor{Beige_ST}{RGB}{255,250,200}
\definecolor{Green_ST}{RGB}{60,180,75}
\definecolor{Lime_ST}{RGB}{210,245,60}
\definecolor{Mint_ST}{RGB}{170,255,195}
\definecolor{Teal_ST}{RGB}{0,128,128}
\definecolor{Navy_ST}{RGB}{0,0,128}
\definecolor{Blue_ST}{RGB}{0,130,200}
\definecolor{MedBlue_ST}{RGB}{0,150,220}
\definecolor{Cyan_ST}{RGB}{70,240,240}
\definecolor{Purple_ST}{RGB}{145,30,180}
\definecolor{Lavender_ST}{RGB}{220,190,255}
\definecolor{DarkLavender_ST}{RGB}{190,160,225}
\definecolor{DarkLavender_ST2}{RGB}{140,110,175}
\definecolor{Magenta_ST}{RGB}{240,50,230}
\definecolor{Grey_ST}{RGB}{128,128,128}

\definecolor{Red_P1CB}{RGB}{251,180,174}
\definecolor{Blue_P1CB}{RGB}{179,205,227}
\definecolor{Green_P1CB}{RGB}{204,235,197}
\definecolor{Lavender_P1CB}{RGB}{222,203,228}
\definecolor{Orange_P1CB}{RGB}{254,217,166}
\definecolor{Yellow_P1CB}{RGB}{255,255,204}
\definecolor{Brown_P1CB}{RGB}{229,216,189}
\definecolor{Pink_P1CB}{RGB}{253,218,236}
\definecolor{Wine_SpeCB}{RGB}{158,1,66}
\definecolor{Red_SpeCB}{RGB}{213,62,79}
\definecolor{Tomato_SpeCB}{RGB}{244,109,67}
\definecolor{Khaki_SpeCB}{RGB}{254,224,139}
\definecolor{Lemon_SpeCB}{RGB}{255,255,191}
\definecolor{Lime_SpeCB}{RGB}{230,245,152}
\definecolor{Green_SpeCB}{RGB}{171,221,164}
\definecolor{AquaMarine_SpeCB}{RGB}{102,194,165}
\definecolor{Blue_SpeCB}{RGB}{50,136,189}
\definecolor{Purple_SpeCB}{RGB}{94,79,162}

\definecolor{NewGreen}{RGB}{19,137,89}
\definecolor{NewBlue}{RGB}{23,146,164}
\newcommand{\Randomcolor}{Grey_ST}
\newcommand{\LECGcolor}{Blue_SpeCB}
\newcommand{\LLRmThreecolor}{Red_SpeCB}
\newcommand{\LLSFcolor}{gold}
\newcommand{\LLRmTwocolor}{NewGreen}
\newcommand{\LLRmOnecolor}{Purple_SpeCB}
\newcommand{\Allcolor}{black}

\def\BibTeX{{\rm B\kern-.05em{\sc i\kern-.025em b}\kern-.08em
    T\kern-.1667em\lower.7ex\hbox{E}\kern-.125emX}}

\begin{document}

\title{Study of Adaptive LLR-based AP selection for Grant-Free Random Access in Cell-Free Networks\vspace{-0.05em}\\
\thanks{This work was supported by the Conselho Nacional de Desenvolvimento Cient\'{i}fico e Tecnol\'{o}gico (CNPq).}
}
\author{\IEEEauthorblockN{Roberto B. Di Renna \textit{Member}, \textit{IEEE} and Rodrigo C. de Lamare, \textit{Senior Member}, \textit{IEEE}}
\IEEEauthorblockA{Center for Telecommunications Studies (CETUC)\\
Pontifical Catholic University of Rio de Janeiro, RJ, Brazil\\
Email: delamare@puc-rio.br \vspace{-0.05em}}
}

\maketitle

\begin{abstract}
    This paper presents an iterative detection and decoding scheme along with an adaptive strategy to improve the selection of access points (APs) in a grant-free uplink cell-free scenario. With the requirement for the APs to have low-computational power in mind, we introduce a low-complexity scheme for local activity and data detection. At the central processing unit (CPU) level, we propose an adaptive technique based on local log-likelihood ratios (LLRs) to select the list of APs that should be considered for each device. Simulation results show that the proposed LLRs-based APs selection scheme outperforms the existing techniques in the literature in terms of bit error rate (BER) while requiring comparable fronthaul load. \vspace{1em}
\end{abstract}

\begin{IEEEkeywords}
Cell-free, mMTC, iterative detection and decoding, grant-free random access.
\end{IEEEkeywords}

\section{Introduction}
    %
    The rapid expansion of  internet-of-things (IoT) solutions has driven a great deal of interest in machine-type communications (MTC) for the next generation of wireless networks. The deployment of IoT applications in different industries require a reliable signal coverage and, given that the actual cellular-based system have its drawbacks as inter-cell interference, new solutions have been advocated in recent years~\cite{DiRennaAccess2020},\cite{IAkyildizAccess2020}. In order to support more users with uniform performance, the cell-free massive multiple-input multiple-output (MIMO) concept recently emerged \cite{mmimo,wence}. The essence is to support a massive number of devices, in the same time-frequency resource, geographically distributing a substantial number of access points (AP)~\cite{HNgoTWC2017}. The system is coordinated by a central processing unit (CPU) that is connected via a fronthaul network to each AP. Thus, in this system there are actually no cell boundaries~\cite{EBjornsonTWC2020}. In theory, the system can achieve excellent gains, but in a situation that each device should be served by all APs in the network~\cite{EBjornsonTCom2020}. Nevertheless, this is not scalable due to huge fronthaul signalling and computational cost requirements. In a scenario where a huge number of machine-type devices (MTCDs) are energy-constrained and transmits small packets~\cite{PopovskiAccess2018}, APs selection schemes performed by the CPU are a practical solution to minimize the network and fronthaul signalling ~\cite{TNguyenAccess2018,rscf}.
	
Previous works that aim to study the APs selection \cite{tds} problem are based on channel statistics. The work in~\cite{HNgoTGCN2018} proposes an APs selection technique based on largest large-scale fading (LLSF) and the geographic distance between APs and UEs.
	Using the signal-to-interference-plus-noise ratio (SINR) as a metric, the work in~\cite{HDaoAccess2020} improves the technique in~\cite{HNgoTGCN2018}. Using an LLSF-based APs selection technique, the work in~\cite{CDAndreaLComm2021} proposes to add another step in each AP processing and send local LLRs on the fronthaul link instead of sending the soft-detected symbols. There are also solutions based on graph neural networks~\cite{VRanasingheGlobecom2021}, sum-rate optimization~\cite{TVuICC2020} and power allocation problems to increase the spectral efficiency~\cite{TVanChienTWC2020, GDongTVT2019, RWangAccess2021, VPalharesTVT2021}. However, these studies do not consider the mMTC scenario.
    %
    In the literature there are few works on mMTC in the cell-free MIMO scenario, but not necessarily propose an APs selection technique. The authors of~\cite{UKGanesanTCom2021} propose activity detection techniques while the works in~\cite{JDingCSCN2021} and~\cite{AMishraCommLet2022} focus on the preamble collision problem. Although the full mMTC scenario is not covered, the work in~\cite{ZWangWCSP2020} studies the effects of using non-orthogonal multiple access in cell-free MIMO, and proposes an APs selection technique based on a joint power optimization strategy. 
    
    %
    This is one of the first works that study mMTC in the cell-free scenario and the APs selection problem. In particular, we propose an iterative detection and decoding (IDD) scheme along an adaptive LLR-based APs selection technique for grant-free random access (GFRA) in cell-free Massive MIMO. Since one of the goals of the cell-free concept is to distribute a huge number of APs with limited resources, we incorporate a regularized algorithm of our previous work~\cite{DiRennaWCL2019} to perform the local activity and data detection. In our cell-free mMTC framework, each AP gives support to all devices locally considered as active. Each AP sends via fronthaul to the CPU the list of supported devices, the received signal and their detected symbols. These are then gathered at the CPU where they are linearly processed to perform joint detection. In this sense we propose an IDD scheme along with an adaptive APs selection technique that uses the soft detected symbols to compute bit log-likelihood ratios (LLR) and update the APs selection list for each device. Simulation results show that including the local LLR computation on the subset APs decision procedure benefits the system performance in terms of BER. Moreover, we also study the fronthaul signalling load and the computational cost required.
    
    This paper is structured as follows. Section II introduces the system model. The problem of decentralized detection, the regularized filter and the activity detection procedure are presented in Section III. Section IV presents the proposed IDD scheme, explaining the LLR computation and APs selection technique. In Section V we discuss the required computational cost and the fronthaul load. Numerical results are in Section VI, whereas the conclusions are drawn in Section VII.

\section{System model}
    We consider a GFRA uplink cell-free mMTC system that consists of $L$ APs each equipped with $N$ antennas which cover an area of $K$ single-antenna devices, as shown in Fig.~\ref{fig:sysmod}(a). The data decoding is performed in the nearest CPU with the information provided by the APs via fronthaul links. In mMTC scenarios, the number of MTCDs $K$ is larger than that of antennas $N$ at each AP such that they consist of underdetermined systems. However, the data symbols can be detected as their vector representation has a sparse structure as the rows corresponding to the inactive users are zero. Assuming that $L$ and $K$ are large and $N L \gg K$~\cite{HNgoTWC2017}, the scenario fits in the massive MIMO framework as it permits coherent joint reception of the signals in the coverage area. 
    
    \subsection{GFRA transmission}
    In the GFRA scenario, when a MTCD is active, it is allowed to transmit a frame at the beginning of any frame interval. Each frame is divided into $\tau_c = \tau_p + \tau_u$, where $\tau_p$ is the number of pilots and $\tau_u$ the data. Since the mMTC traffic is intermittent, the model results in sparse systems, where we designate the Boolean variable $\alpha_k = 1$ that indicates the activity of the $k-$th device in the observation window and $\alpha_k = 0$, otherwise. The transmitted data of the $k-$th MTCD, $\bx_k$, is composed by symbols from a regular modulation scheme, as quadrature phase-shift keying (QPSK). Due to the huge number of devices, the CPU periodically distributes independent non-orthogonal pilot sequences to the APs broadcast to the devices. When a MTCD is active, it randomly chooses one of those sequences given by $\boldsymbol{\phi}_k = \exp\left(j\pi\boldsymbol{\nu}\right)$, where each element of vector $\boldsymbol{\nu} \in \mathbb{R}^{\tau_p}$ is drawn uniformly at random in $[-1,1]$. Despite the intermittent pattern of transmissions, each device should wait, at least, for the guard period interval to transmit again.
    
    
    \subsection{Channel model}
    We consider the block fading model to describe the channel, where it is constant over a transmission frame duration $\tau_c$ and changes independently from each coherence block. In our model we take into account the geography of the APs to model the large-scale fading, which includes geometric pathloss, shadowing, antenna gains and spatial channel correlation~\cite{AGoldsmithJSAC2003}. Therefore, for the $k-$th MTCD and the $l-$th AP, the channels are represented as independent correlated Rayleigh fading realizations given by circularly symmetric complex Gaussian distribution with zero mean and spatial correlation matrix $\bR_{lk} \in \mathbb{C}^{N\times N}$, that is, $\dot{\bh}_{lk} \sim \mathcal{N}_{\mathbb{C}}\left(\mathbf{0},\bR_{lk}\right)$, where the large-scale fading parameter is described as $\beta_{lk} = \text{tr}\left(\bR_{lk}\right)/N$. Due to the scenario, we also assume that the channel vectors of each AP are independent and identically distributed (i.i.d.), thus $\mathbb{E}\{\dot{\bh}_{kn}(\dot{\bh}_{lk})^{\text{H}}\}=\mathbf{0}$ and that the channels of each MTCD are also i.i.d.~\cite{EBjornsonTCom2020,EBjornsonTWC2020}.
    %
    In this context, for the $l-$th AP, the received signal $\mathbf{y}_l \in \mathbb{C}^{N \times 1}$ is given by
    \begin{equation}\label{eq:sysmod}		
        \mathbf{y}_l\t =\left\{ 
    	\begin{array}{ll}
    		 \mathbf{H}_l \, \boldsymbol{\phi}\t + \mathbf{v}_l\t,& \text{if}\hspace{10pt} 1 \leq t \leq \tau_p  \\ 
             \mathbf{H}_l \, \mathbf{x}\hspace{1pt}\t + \mathbf{v}_l\t ,& \text{if}\hspace{10pt}  \tau_p  < t \leq \tau_c
    	\end{array}\right.
    \end{equation}
    
    \noindent where the channel matrix $\bH_l \in \mathbb{C}^{N \times K}$ is sparse, as $\bH_l = \dot{\bH_l}\, \text{diag}{(\boldsymbol{\alpha})}$ and the transmitted signals are represented by the symbol interval indicator $t$, in which $\bphi$ and $\bx \in \mathbb{C}^{K \times 1}$. The noise vector with i.i.d. $\mathcal{N}_c\left(0, \sigma^2_v\right)$ entries is given by $\bv_l \in \mathbb{C}^{N \times 1}$, wherein $\sigma^2_v$ is the noise power. 

    \begin{figure*}[t]
        \begin{minipage}[b]{0.4\linewidth}
            \hspace{0.05cm}
            \includegraphics[scale=0.88]{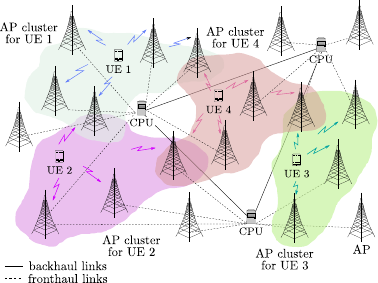}
            \centerline{\footnotesize (a)}\medskip
        \end{minipage}\hfill
        \begin{minipage}[b]{0.5\linewidth}
            \includegraphics[scale=0.88]{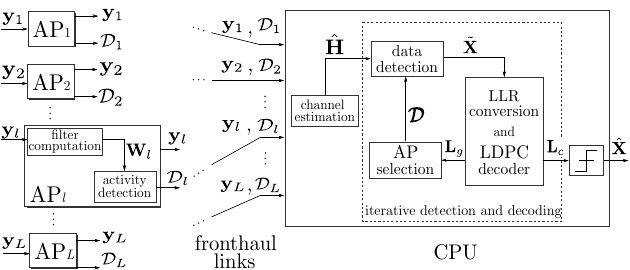}
            \centerline{\footnotesize (b)}\medskip
        \end{minipage}        
        \vspace{-0.4cm}
		\caption{(a) Representation of the cell-free architecture and (b) block diagram of one CPU connection. Each AP computes local filters, performs the activity and data detection and transmits via fronthaul to the CPU. The CPU realizes iteratively the detection and decoding procedure updating the APs selection list $\boldsymbol{\mathcal{D}}$ at every new iteration.}
		\label{fig:sysmod}%
    \end{figure*} 

\section{Local activity and data detection}
Since one of the premises of cell-free networks is to distribute a substantial number of APs with reduced processing power, a complex detector in each AP is impracticable. Thus, in order to use a low-complexity solution and avoid possible throughput and latency issues, we adapt the method in~\cite{DiRennaWCL2019} to perform the local joint activity and data detection. In the following we describe the basic idea of the Activity-Aware Recursive Least Squares with Decision-Feedback (AA-RLS-DF) algorithm and refer to the paper~\cite{DiRennaWCL2019} for further details. We remark that other interference cancellation techniques can be adopted \cite{jidf,spa,mfsic,dfcc,did,bfidd,1bitidd,aaidd,dynovs,detmtc,comp,msgamp,msgamp,msgamp2}.

%
Unlike the literature, where there is an assumption that each AP have knowledge of the spatial correlation matrices, in GFRA the MTCDs simply transmit and this information is not available~\cite{EBjornsonTCom2020}. Thus, in our approach, each AP does not need to estimate the channels, the work of joint activity and data detection is performed by the AA-RLS-DF algorithm. As described in Fig.\ref{fig:sysmod}(b), each AP computes the receive filter and performs the main task.

The AA-RLS-DF algorithm employs an $l_0$-norm regularization with decision feedback that reduces the detection error propagation. As the transmission slots are separated by pilots and data, AA-RLS-DF has two modes of operation, training mode ($1 \leq t \leq \tau_p$) and decision-directed mode ($\tau_p < t \leq \tau_c$). Although the same scheme is used for both cases, in the training mode the focus is to use the pilots to update the RLS algorithm, while the decision-directed mode uses the filter to detect the received data symbols. AA-RLS-DF detects each symbol at a time, per layer. The detection order is updated at each new layer, using the least squares estimation (LSE) criterion. The adaptive receive filter can be decomposed into feedforward and feedback filters. The feedforward one is updated at every new received vector by the $l_0$-norm regularized RLS algorithm. The feedback filter is a component that is concatenated to the feedforward filter in order to cancel the interference of the previously detected symbols.
%
In order to exploit the sparse activity of MTCDs and compute the receive filter parameters without requiring explicit channel estimation, we included in the algorithm an $l_0$-norm regularization
that minimizes the cost function. Approximating the value of the $l_0$-norm~\cite{Bradley98}, taking the partial derivatives for all entries $t$ of the coefficient vector $\mathbf{w}_{k}\t$, setting the results to zero and suppressing the AP index $l$, yields
    \begin{equation}\label{eq:recursive_rls_l0_app}
		\mathbf{w}_{k}\t =\mathbf{w}_{k}\left[t-1\right] + \mathbf{g}\t \varepsilon_k\ast\t - \gamma \, \epsilon\, \text{sgn}\left(w_{kn}\t\right)f_\epsilon\left(w_{kn}\t\right),
	\end{equation}
	
\noindent  where $\mathbf{g}\t$ is the gain vector and $\text{sgn}\left(\cdot\right)$ is a component-wise sign function defined as	$\text{sgn}\left(w_{kn}\t\right) = w_{kn}\t/|w_{kn}\t|\, \forall \, w_{kn}\t \neq 0$ and $\text{sgn}\left(w_{kn}\t\right) =0$, otherwise. The attraction to zero of small coefficients is imposed by the function $f_\epsilon\left(w_{kn}\t\right)$ and the parameters $\epsilon$ and $\gamma$.

		

%
Once the AA-RLS-DF algorithm ends, the list of MTCDs judged as active by the $l-$th AP, $\boldsymbol{\mathcal{D}}_l$, is a binary vector formed based on the non-zero column norms of the filter matrix $\mathbf{W} \in \mathbb{C}^{N \times K}$. Thus, each AP transmits $\boldsymbol{\mathcal{D}}_l$ and $\bY_l$ via fronthaul to the CPU.


\section{CPU data detection}
In this section we present the CPU data detection, explaining the LLR computation and the APs selection technique. Specifically, each AP in the uplink transmits the received signals, the activity detection list and the local soft MIMO detection of the data to the CPU for final decoding.
We can divide our scheme in two steps. In the first step, the CPU computes each AP LLRs and evaluates reliability of the whole frame per MTCD. If the mean of the LLRs of the $k-$th MTCD computed by the $l-$th AP is higher than a threshold, the LLRs of the $l-$th AP participates in the joint detection and decoding with the other APs that gives support to the $k-$th MTCD. After updating the list $\boldsymbol{\mathcal{D}}$, in the second step the LLRs of the $k-$th MTCD are recomputed, but now with the information of all APs that gives support to it, i.e., $\mathcal{D}_{lk} = 1$.

\subsection{Channel statistics}
Under the assumption that the CPU has substantially more processing resources than the APs, we consider the MMSE filter in the IDD procedure. As in this scenario we have spatial correlated channels and also considering the estimation error, we take it into account in the filter computation, following the literature~\cite{EBjornsonTCom2020,JWangSPL2009}. As in this part of the derivation we are using complete parts of the matrices, the time index is suppressed to avoid a heavy notation. At this point, the CPU has the information of each APs activity detection and under the assumption that each device has its own identification~\cite{DiRennaAccess2020}, the channel can be estimated as given by
    \begin{equation}\label{eq:channel_est}
        \hat{\bH}_l = \frac{1}{E_p \tau_p}\, \bY_{l}^{p} \, \bPhi^\text{H} \boldsymbol{\mathcal{D}}_l^\text{T} = \bH_l + \Delta\bH_l,
    \end{equation}

\noindent where $E_p$ is the energy of each pilot symbol and $\bY_{l}^{p}$ is the part of the received matrix that comprises the transmitted pilot symbols $\bPhi$, $(1 \leq t \leq \tau_p)$. The estimation error of $\bH_l$ is given by 
    \begin{equation}\label{eq:EstErr}
        \Delta\bH_l = \frac{1}{E_p \tau_p}\, \bV_l^p \bPhi^\text{H}\boldsymbol{\mathcal{D}}_l^\text{T},
    \end{equation}

\noindent where the noise matrix $\bV_l^p$ follows the same idea as $\bY_{l}^{p}$. Based on~(\ref{eq:EstErr}),  one can see that $\Delta\bH_l$ is uncorrelated with $\bH_l$ and has i.i.d. $\mathcal{C}\mathcal{N}\left(0,\sigma^2_{\Delta h}\right)$ entries. Thus, $\sigma^2_{\Delta h}$ can be given by $ \sigma^2_{\Delta h} = \nicefrac{\sigma^2_{v}}{E_p \tau_p}$.
%

Using the assumptions above, we can also conclude that $\hat{\bH}_l$ is a complex Gaussian random matrix with zero mean. Using properties of complex Gaussian random vectors~\cite{SMKay1993} and with some manipulations, we have the following expressions:
    \begin{align}
        \mathbb{E}\left[\hat{\bh}_{lk} \hat{\bh}_{lk}^\text{H}\right] & = \hat{\bR}_{lk} + \sigma^2_{\Delta h} \mathbf{I}_{N},\\ \label{eq:MeanDeltah}
        \mathbb{E}\left[\Delta h_{lk} | \hat{\bh}_{lk}\right] & = \sigma^2_{\Delta h}\left(\hat{\bR}_{lk} + \sigma^2_{\Delta h} \mathbf{I}_{N}\right)^{-1} \hat{\bh}_{lk},\\ \label{eq:CovDeltah}
        \text{Cov}\left[\Delta \bh_{lk} \Delta \bh_{lk}^{\text{H}} | \hat{\bh}_{lk}\right] & = \sigma^2_{\Delta h_l} \mathbf{I}_{N} - \sigma^4_{\Delta h_l}\left(\hat{\bR}_{lk} + \sigma^2_{\Delta h_l} \mathbf{I}_{N}\right)^{-1}
    \end{align}

\noindent where $\text{Cov}\left[\cdot\right]$ denotes covariance operation and $\bR$ can be estimated as $\hat{\bR}_{lk}=  \nicefrac{1}{K^\text{act}_l}\, \hat{\bH}_l\hat{\bH}_l^\text{H} - \sigma^2_{\Delta h_l} \mathbf{I}_{N}$ where $K^\text{act}_l$ is the number of MTCDs detected as active for the $l-$th AP.
\subsection{Computation of MMSE filter}
With the channel statistics presented, we proceed to the MMSE filter computation. Considering the estimation error and the data frame part, (\ref{eq:sysmod}) can be rewritten as given by
    \begin{align}
        \bY_l = \left(\hat{\bH}_l - \Delta\bH_l\right) \bX_l + \bV_l. 
        \vspace{-0.1cm}
    \end{align}
Then, conditioned on $\hat{\bH}_l$, the MMSE filter is given by~\cite{SMKay1993} 
    \begin{equation}\label{eq:MMSEfilter}
        \bW_l = \mathbb{E}\left[\bX\bY_l^\text{H}|\hat{\bH}_l\right]\left\{\mathbb{E}\left[\bY_l\bY_l^\text{H}|\hat{\bH}_l\right]\right\}^{-1},
        \vspace{-0.1cm}
    \end{equation} 
\noindent where
    \begin{align}\label{eq:MeanYYgivenHhat}
        & \mathbb{E}\left[\bY_l\bY_l^\text{H}|\hat{\bH}_l\right] =  E_s\left(\hat{\bH}_l\hat{\bH}_l^\text{H} - \hat{\bH}_l\mathbb{E}\left[\Delta\bH_l^\text{H} | \hat{\bH}_l\right]\right.& \\ \nonumber
        & \hspace{1cm} \left. - \mathbb{E}\left[\Delta {\bH}_l | \hat{\bH}_l\right] \hat{\bH}_l^\text{H} + \mathbb{E}\left[\Delta {\bH}_l \Delta {\bH}_l^\text{H} | \hat{\bH}_l\right]\right) + \sigma^2_v \mathbf{I}_N.&
    \end{align}

\newpage 
Considering $E_s$ as the symbol energy of the active devices and using the statistics of the previous subsection, we have
\vspace{-0.1cm}
    \begin{align}\label{eq:MeanDeltaDeltaH}
        &\mathbb{E}\left[\Delta\bH_{l} \Delta \bH^\text{H}_{l} | \hat{\bH}_{l}\right] = {\textstyle \sum_{k=1}^{K^\text{act}_l}}\, \mathbb{E}\left[\Delta \bh_{lk} \Delta \bh_{lk}^\text{H} | \hat{\bh}_{lk}\right],\, \text{as}& \\ \label{eq:CovDeltaDeltah}
        &\mathbb{E}\left[\Delta \bh_{lk} \Delta \bh_{lk}^\text{H} | \hat{\bh}_{lk}\right] =&\\ \nonumber 
        &\hspace{0.5cm} \text{Cov}\left[\Delta \bh_{lk} \Delta \bh_{lk}^\text{H} | \hat{\bh}_{lk}\right] + \mathbb{E}\left[\Delta \bh_{lk} | \hat{\bh}_{lk}\right] \mathbb{E}\left[\Delta \bh_{lk}^\text{H} | \hat{\bh}_{lk}\right]&
    \end{align}

\noindent Including (\ref{eq:MeanDeltah}) and (\ref{eq:CovDeltah}) into (\ref{eq:CovDeltaDeltah}) yields
\begin{align}
        &\mathbb{E}\left[\Delta\bh_{lk} \Delta \bh^\text{H}_{lk} | \hat{\bh}_{lk}\right] =&  \\ \nonumber
        & \hspace{2cm} \sigma^2_{\Delta h} \mathbf{I}_N + \sigma^4_{\Delta h} \left(\hat{\bh}_{lk} \hat{\bh}_{lk}^\text{H} - 1\right) \left(\bR_{lk} + \sigma^2_{\Delta h} \mathbf{I}_N\right)^{-1}& 
\end{align}

Defining $\boldsymbol{\Theta}_{lk} = (\hat{\bR}_{lk} + \sigma^2_{\Delta h} \mathbf{I}_{N})^{-1}$ and $\boldsymbol{\Omega}_l = [\boldsymbol{\Omega}_{l1}, \dots, \boldsymbol{\Omega}_{lK}]$, where $\boldsymbol{\Omega}_{lk} =  \boldsymbol{\Theta}_{lk} \hat{\bh}_{lk}$  and inserting (\ref{eq:CovDeltaDeltah}) into (\ref{eq:MeanDeltaDeltaH}), we have
    \begin{align}\label{eq:MeanDeltaDeltaHhat}
       &\mathbb{E}\left[\Delta \bH_l \Delta \bH_l^\text{H}\right | \hat{\bH}_l] =& \\ \nonumber
       &\hspace{1.5cm} K^\text{act}_l \sigma^2_{\Delta h}\mathbf{I}_N + \sigma^4_{\Delta h} \left(\boldsymbol{\Omega}_l \boldsymbol{\Omega}^\text{H}_l - \sum^{K^\text{act}_l}_{k=1} \boldsymbol{\Theta}_{lk}\right).&       
    \end{align}

Similarly, we can also get $\mathbb{E}\left[\Delta \bH\right | \hat{\bH}] = \sigma^2_{\Delta h} \bOmega$ and inserting it and (\ref{eq:MeanDeltaDeltaHhat}) into (\ref{eq:MeanYYgivenHhat}), we have
    \begin{align}\nonumber
        & \mathbb{E}\left[\bY_l\bY_l^\text{H}|\hat{\bH_l}\right] =  E_s\left(\hat{\bH}_l - \sigma^2_{\Delta h}\bOmega_l\right)\left(\hat{\bH}_l - \sigma^2_{\Delta h}\bOmega_l\right)^\text{H}& \\ \label{eq:MeanYYgivenHhat2}
        &\hspace{0.5cm} + E_s \sigma^2_{\Delta h} \left(K^\text{act}_l \mathbf{I}_N - \sigma^2_{\Delta h} \sum^{K^\text{act}_l}_{k=1} \boldsymbol{\Theta}_{lk}\right) + \sigma^2_v \mathbf{I}_N.&
    \end{align}

To conclude the filter computation in~(\ref{eq:MMSEfilter}), we have
    \begin{equation}
        \mathbb{E}\left[\bX \bY^\text{H}_l | \hat{\bH}_l\right] = E_s \left(\hat{\bH}^\text{H}_l - \sigma^2_{\Delta h} \bOmega^\text{H}_l\right).
    \end{equation}

With the local filters computed, the CPU computes the local LLRs and applies the LLR-based APs selection.

\subsection{LLR computation}
    The proposed structure, represented in Fig.\ref{fig:sysmod}(b), consists of the exchange of extrinsic information of the channel decoder and the intrinsic information calculated by the soft MIMO detector. The information exchanged between the detector and decoder is done in an iterative fashion until a maximum number of iterations is reached~\cite{DiRennaTCom2020,XWang1999}. In the proposed IDD scheme we employ low-density parity-check (LDPC) codes \cite{richardson,memd}.
    
    As the IDD structure exchanges information per code bit, we have that the extrinsic information computed by the MMSE detector is the difference of the soft-input and soft-output LLR values. In the $l-$AP, let $b_k(i)$ represent the $i-$th bit of the modulated symbol $x_k$, transmitted by the $k-$th UE. Considering $M_c$ as the modulation order ($i \in \{1,\dots, M_c\}$), the extrinsic LLR value of the estimated bit ($b_k(i)$) is
    \vspace{-0.1cm}
    \begin{align} \nonumber
        &\hspace{-0.12cm} \text{L\small{g}}_{lk}\left(b_{k}(i)\right) = \log \frac{P\left(b_{k}(i) = 1 | \hat{x}_{lk}\right)}{P\left(b_{k}(i) = 0 | \hat{x}_{lk}\right)} - \log \frac{P\left(b_{k}(i) = 1\right)}{P\left(b_{k}(i) = 0\right)} \\ \label{eq:LLR}
        &\hspace{0.6cm} = \log \frac{\sum_{x_k\in\mathcal{A}_i^{1}} P\left(\hat{x}_{lk}|x_k\right) P\left(x_k\right)}{\sum_{x_k\in\mathcal{A}_i^{0}} P\left(\hat{x}_{lk}|x_k\right) P\left(x_k\right)} - \text{L\small{c}}_{lk}\left(b_{k}(i)\right)\!,
    \end{align}     
    %
 
    \noindent where $\text{L\small{c}}_{lk}\left(b_{k}(i)\right)$ is the extrinsic information of $b_k(i)$ computed by the LDPC decoder in the previous turbo iteration. $\mathcal{A}_i^{1}$ and $\mathcal{A}_i^{0}$ represents the set of $2^{M_c-1}$ hypotheses for which the $i-$th bit is 1 or 0, respectively. The soft information $\tilde{x}_{lk}$ provided by the MIMO detector is given by 
    \begin{equation}\label{eq:SoftDet}
        \tilde{x}_{lk} = \bw_{lk}^\text{H}\, \by_l,\hspace{1cm} \forall \, \mathcal{D}_{lk} =1.
    \end{equation}    
    
    The \textit{a priori} probability $P(x_k)$ in (\ref{eq:LLR}) is given by 
    \begin{equation}\label{eq:Px}
        P(x_k) = \prod^{M_c}_{i=1} \left[1 + \exp{\left(-\overline{b}_k(i)\, \text{L\small{c}}_{lk}\left(b_{k}(i)\right)\right)} \right]^{-1}    
    \end{equation}

    \noindent where $\overline{b}_k(i)$ denotes the state \footnote{Ex.: for QPSK, if $x_k =\nicefrac{\sqrt{2}}{2}(1-j)$, we have $b_{k}(1)=1$, $b_k(2)=0$, $\overline{b}_k(1)=+1$ and $\overline{b}_k(2)=-1$.} of the $i-$th bit of the symbol of the $k-$th user~\cite{XWang1999}. Finally, the likelihood function $ P\left(\tilde{x}_{lk}|x_k\right)$ is approximated by
    \begin{equation}\label{eq:likelihood}
        P\left(\tilde{x}_{lk}|x_k\right) \simeq \frac{1}{\pi \eta^2_{lk}} \exp{\left(-\frac{1}{\eta^2_{lk}} |\tilde{x}_{lk} - \mu_{lk} x_k|^2\right)}.
    \end{equation}
    
    Since the soft symbols are degenerated versions of the constellation list, the MMSE filter outputs are neither Gaussian nor i.i.d., which requires an approximation to compute the means $\mu$ and variances $\eta$ in (\ref{eq:likelihood}). Thus, we approximate $\tilde{x}_{lk}$ by the output of an equivalent AWGN channel with $\tilde{x}_{lk} \approx \mu_{lk} x_k + z_k$, where
    \begin{align} \nonumber
        \mu_{lk} &= \mathbb{E}\left[\bw_{lk}\left(\hat{\bh}_{lk} - \Delta\bh_{lk}\right) \big| \hat{\bH}\right] - \boldsymbol{\mu}_{lk}^2 E_s& \\
        & = \bw_{lk}^\text{H}\left(\hat{\bh}_{lk} - \sigma^2_{\Delta h} \bOmega_{lk}\right)&
    \end{align}
    
    \noindent and $z_k \sim \mathcal{N}_c(0,\eta^2_k)$, where
    \begin{align} \label{eq:varLLR}
        \hspace{-0.4cm}\eta^2_{lk} = E_s \left(\hat{\bh}^\text{H}_{lk} \left(\mathbf{I}_{N} - \sigma^2_{\Delta h} \boldsymbol{\Theta}_{lk}\right) - \mu_{lk}^2\right).
    \end{align}    

    Since all the local LLRs are computed, the CPU can proceed to the LDPC decoder, as depicted in Fig.\ref{fig:sysmod}(b). The local LLRs go to an evaluation step that determines which one will join the final detection, i.e., be part of the list of the APs that participate in the detection of an specific MTCD. Improvements in the decoding can be obtained by reweighting \cite{vfap} or scheduling \cite{kaids} approaches.


\subsection{LLR-based APs selection}
    The main idea is, at every new iteration, evaluate the reliability of the local LLRs of each supported MTCD in AP $l$ to a threshold and, in case of success, include the corresponding estimated channel, filter and received signal in the joint detection and decoding. Thus, the proposed APs selection algorithm is described in the following steps:
    \begin{enumerate}
        \item After the GFRA procedure, each AP computes the AA-RLS-DF filters with (\ref{eq:recursive_rls_l0_app}) and performs local activity detection, thus creating its first version of $\mathcal{D}_l$. In our solution, each AP includes in $\mathcal{D}_l$ the MTCDs index with filter norm different from 0, that is,
        \begin{equation}
            \text{if } \|\bw_{lk}\| \neq 0, \mathcal{D}_{lk} = 1.
        \end{equation}

        \item Each AP transmits via fronthaul to the CPU the support vector $\mathcal{D}_l$ and the received signal $\bY_l$;
        
        \item The channels and the spatial correlation matrices are estimated by the CPU with (\ref{eq:channel_est})-(\ref{eq:CovDeltah});
        %
        %
        %
        \item Computing the filters with (\ref{eq:MMSEfilter}), the CPU calculates the local LLRs of each symbol of the whole data frame by the IDD scheme with~(\ref{eq:LLR})-(\ref{eq:varLLR}); 
        \item The CPU decides if the detected soft-symbol of the $k-$th device by the $l-$th AP should be considered in the joint detection based on its reliability. If the mean of the absolute value of the LLRs of the transmitted frame of the $k-$th MTCD is larger than a predefined threshold value $T_h$, the information processed by the $l-$th AP regarding the $k-$th MTCD is included in the list, as given by:
        \begin{equation}
            \text{if } \mathbb{E}\left\{\mathbf{L}\text{\small g}_{lk}\right\} \geq T_h,   k \in \boldsymbol{\mathcal{D}}_l.
        \end{equation}        

        %
        \item By then, each $\boldsymbol{\mathcal{D}}_l$, channel estimation and filters in~(\ref{eq:MMSEfilter}) are updated for the supported devices. 

        \item With the supporting matrix $\boldsymbol{\mathcal{D}}$ updated, for the CPU to perform the joint detection and decoding, the received signals (data frame part), the estimated channels and the filters computed in the previous section are concatenated as $\bY = \left[\bY_1, \cdots, \bY_l, \cdots, \bY_L\right]^\text{T} \in \mathbb{C}^{N L \times \tau_u},$ $\bH = \left[\bH_1, \cdots, \bH_l, \cdots, \bH_L\right]^\text{T}$ and $\bW = \left[\bW_1, \cdots, \bW_l, \cdots, \bW_L\right]^\text{T} \in \mathbb{C}^{N L \times K}$.
        %
        
        %
        Therefore, the soft information provided by the MIMO detector in~(\ref{eq:SoftDet}) is now given by
        \begin{equation}
            \tilde{x}_k\t = \left(\bw_k\right)^\text{H}\, \text{diag}\left\{\bD_k\right\} \by\t, \forall \, \tau_p  < t \leq \tau_c,
        \end{equation}
    
        \noindent where $\bD \in \mathbb{C}^{NL \times K}$ is the list of supported MTCDs per AP, detailed in the number $N$ of AP antennas;

        \item The joint LLR computation proceeds as in~(\ref{eq:LLR})-(\ref{eq:varLLR}) and the IDD scheme goes back to step 4 and continues until the number of iterations reaches its maximum.
    \end{enumerate}

    Different values of threshold can be used, but in this work we decided to check the reliability of the LLRs. In a nutshell, if the mean of the absolute value of the LLRs of the transmitted frame of the $k-$UE, is larger than the mean of the frames of all UEs in $\mathcal{D}_l$, the device is supported, as $ T_h = \mathbb{E}\left\{\mathbf{L}\text{\small g}_{l}\right\}$.
        %
\section{Computational cost and Scalability}
    In this section, we discuss the cost and the scalability of the APs selection scheme. Considering the ``Random'' APs selection as an upper bound and ``All APs'' as an ideal case where all active MTCDs can be supported by all APs, we included in the comparison the state-of-the-art LLSF~\cite{EBjornsonTWC2020,SChenJSAC2021,RWangAccess2021} and LECG~\cite{HDaoAccess2020} metrics and the proposed LLR-based, described in the previous section. At this cooperation level, all procedures have the same computational complexity in terms of floating point operations (FLOPS), to the filter computation. The LLR computation part adds $2M_c$ to compute~(\ref{eq:Px}), $2\cdot2^{M_c}$ to compute~(\ref{eq:LLR}) and 4 FLOPS to evaluate~(\ref{eq:likelihood}). Let $|\mathcal{M}_k|$ be the set of APs that serves the $k-$th MTCD, thus, the computational cost of the LLR-based APs selection schemes, per symbol, is given by 
    \begin{align}\nonumber
       &(\nicefrac{K}{2})\left(|\mathcal{M}_k|^2+|\mathcal{M}_k|\right) + (\nicefrac{1}{3})\left(|\mathcal{M}_k|^3 - |\mathcal{M}_k|\right) +\\
       &|\mathcal{M}_k|^2 + 2(M_c + 2^{M_c}) + 4.        
    \end{align} 
    As in our framework the initial APs selection is based on the activity detection, the fronthaul load is analyzed in a situation of the subsequent transmissions. Thus, the CPU should inform the APs the list of MTCDs that should be supported for the next coherence time (if the MTCD is detected as active). Thus, the fronthaul load expression given in terms of the number of complex scalars transmitted, is the same for every scheme, since it depends only on $K$ and $|\mathcal{M}_k|$, as given by~\cite{SChenJSAC2021,EBjornsonTCom2020}
    \begin{equation}
        K|\mathcal{M}_k| + \left(|\mathcal{M}_k|^2 K^2 + K|\mathcal{M}_k|\right)/2.
    \end{equation}

    Fig.~\ref{fig:scal}(a) depicts the fronthaul load of different APs selection schemes under $L = 9$ and $L =25$, which shows that even increasing the network, the hierarchy of the required fronthaul signalling load of every scheme persist. One can see in Fig.~\ref{fig:scal}(b), that the mean number of selected APs of the LLR-based APs selection scheme is considerably lower than the fully connected approach. Besides varying the APs number $L$, this simulation considers $N=4$ antennas per AP and $K = \lfloor L/4 \rfloor$ MTCDs. Regarding our main contribution, the LLR-based, requires more fronthaul load than the approaches initialized by LLSF or LECG, but as we will see in numerical results, the gain in performance is significant.

\section{Simulation results}
    The numerical results of the proposed LLR-based APs selection schemes are presented and discussed, being compared with state-of-the-art solutions in terms of BER. In order to represent the cell-free scenario, we simulated an area of $1\, \text{Km}^2$ where $L$ APs equipped with half-wavelength-spaced uniform linear arrays with $N$ antennas are independently and uniformly distributed giving support to $K$ single-antenna MTCDs. The vertical distance of the APs to the UEs is $10$ m, while the communication bandwidth centered over the carrier frequency of $2$ GHz is $20$ MHz, the power spectral density of the noise is $174$ dBm/Hz, and the noise figure is $5$ dB. Albeit the propagation model in~\cite{3GPPTS36814} is designed for cellular networks, we can adapt it to the cell-free model in order to compute the spatial correlation matrix, as seen in~\cite{EBjornsonTCom2020, CDAndreaLComm2021}. Thus, we have $\beta_{lk} = S_c\cdot(-30.5 -36.7 \log_{10}\left(\nicefrac{\delta_{lk}}{1\, \text{m}}\right) + S_{F_{lk}})$, where $\delta_{lk}$ is the distance between MTCD $k$ and AP $l$ and $S_{F_{lk}} \sim \mathcal{N}_\mathbb{C}\left(0,4^2\right)$ is the shadow fading. Additionally, the spatial correlation is produced using the Gaussian local scattering model, $S_c$, with $15^\circ$ angular standard deviation~\cite{EBjornsonTCom2020}. The frames have $\tau_c = 256$ symbols, where $\tau_u = 128$ is used for uplink data transmission. As the proposed APs selection scheme is based on the LLRs computation, the channel coding considered is LDPC with rate $R_\text{\footnotesize LDPC} = 1/2$ and the modulation scheme is QPSK. The average SNR of the $l-$th AP is given by $\text{SNR}_l = \sum_{i=1}^K \beta_l \eta_i R_\text{\footnotesize LDPC} \left(\nicefrac{1}{\sigma_{v_l}^2}\right)$.

    We compare different APs selection techniques in the mMTC cell-free scenario of $L = 100$, $K = 80$ and $N=4$ where the mean activity of the devices is $10\%$. All solutions use the same channel estimation, activity and data detection procedures, with the same constants chosen as in~\cite{DiRennaWCL2019}, thus differing only in the APs selection technique. One can see in Fig.~\ref{fig:BER_SNR} that the proposed LLR-based metrics outperforms the state-of-the-art LLSF and LECG without compromising the fronthaul load, as depicted in Fig.~\ref{fig:scal}. In order to check the performance of different LLR-based metrics, we tested a variation of the threshold metrics. In order to increase the number of selected APs, we considered $T_{h_2} = T_{h_1}/2$ and $T_{h_3} = T_{h_1}/4$, which naturally improves the efficiency as the diversity order is higher. 

%
    \begin{figure}[t]
        \centering
        \input{fig/BERxSNR.tex}
        \vspace{-0.5cm}
        \caption{BER vs. SNR in dB for the different APs selection schemes.}
        \label{fig:BER_SNR}
    \end{figure}
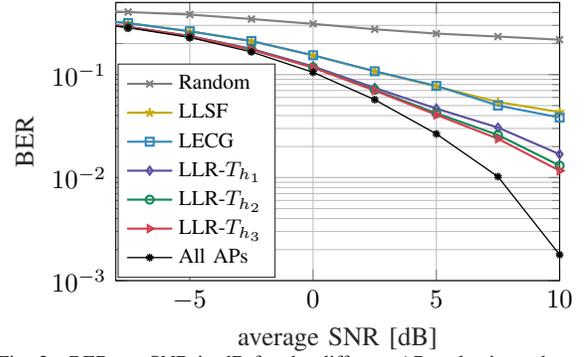    
        \begin{figure}[t]
        \hfill
            \begin{minipage}[b]{0.45\linewidth}
                \centerline{\input{fig/FL_K_L9e25.tex}}
                \centerline{\footnotesize (a)}\medskip
            \end{minipage}
            \hfill
            \begin{minipage}[b]{0.47\linewidth}
                \centerline{\input{fig/Numb_Net.tex}}
                \centerline{\footnotesize (b)}\medskip
            \end{minipage}      
            \vspace{-0.5cm}
    	    \caption{Scalability parameters. (a) Fronthaul load in terms of complex scalars vs. different $K$ values for $N=4$, $L = 9, 25$ and (b) Mean number of used APs per user vs. different network size.}
    	    \label{fig:scal}%
        \end{figure}
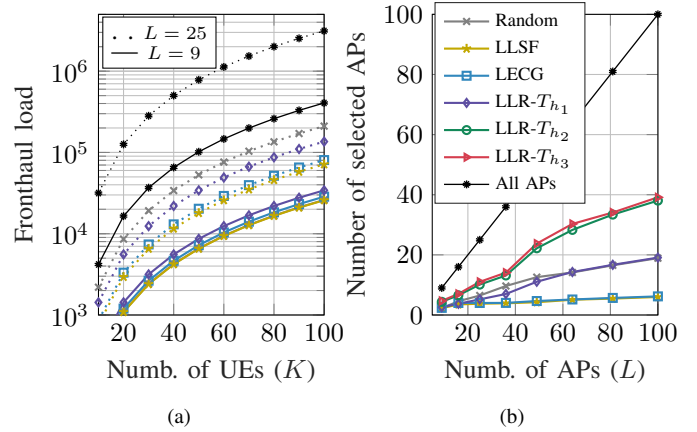     
\section{Conclusions}
    We have studied the APs selection problem in the uplink cell-free mMTC scenario. We have proposed an IDD with LLR-based APs selection scheme that considerably reduces the number of APs connected and thus the fronthaul load. By exploiting the reliability of the locally computed LLRs, the proposed scheme iteratively adapts the list of supported MTCDs of each AP. The results show that the proposed scheme outperforms state-of-the-art approaches in terms of BER with competitive fronthaul load and computational cost.

\bibliographystyle{IEEEbib}
\bibliography{ref_updated}

\end{document}

%% file: fig/BERxSNR.tex
\begin{tikzpicture}
\begin{axis}[%
width=5.9cm,
height=3.7cm,
scale only axis,
xmin=-8,
xmax=10,
xlabel style={font=\color{white!15!black}},
xlabel={average SNR [dB]},
ymode=log,
ymin=0.001,
ymax=0.5,
yminorticks=true,
ylabel style={font=\color{white!15!black}},
ylabel={BER},
axis background/.style={fill=white},
title style={font=\bfseries, align=center},
xmajorgrids,
ymajorgrids,
yminorgrids,
legend style={at={(0.36,0.01)}, font=\footnotesize, anchor=south east, legend cell align=left, align=left, draw=white!15!black},
cycle multi list={Set2-7} 
]

\addplot [color=\Randomcolor, line width=0.8pt, mark=x,mark size = 1.8pt, mark options={solid, \Randomcolor}]
  table[row sep=crcr]{%
    -12.5	0.445029513888889\\
    -10	0.4261875\\
    -7.5	0.404234375\\
    -5	0.381989583333333\\
    -2.5	0.345777777777778\\
    0	0.310559027777778\\
    2.5	0.275213541666667\\
    5	0.249817708333333\\
    7.5	0.233831597222222\\
    10	0.217980902777778\\
    12.5	0.222539930555556\\
    15	0.238902777777778\\
};
\addlegendentry{Random}

\addplot [color=\LLSFcolor, line width=0.8pt, mark=star,mark size = 1.8pt, mark options={solid, \LLSFcolor}]
  table[row sep=crcr]{%
    -12.5	0.391206597222222\\
    -10	0.356107638888889\\
    -7.5	0.315534722222222\\
    -5	0.262677083333333\\
    -2.5	0.210875\\
    0	0.153684027777778\\
    2.5	0.107626736111111\\
    5	0.0765902777777778\\
    7.5	0.0542083333333333\\
    10	0.04321875\\
    12.5	0.0431440972222222\\
    15	0.0426979166666667\\
};
\addlegendentry{LLSF}

\addplot [color=\LECGcolor, line width=0.8pt, mark=square,mark size = 1.5pt, mark options={solid, \LECGcolor}]
  table[row sep=crcr]{%
    -12.5	0.392435763888889\\
    -10	0.356217013888889\\
    -7.5	0.315883680555556\\
    -5	0.263133680555556\\
    -2.5	0.211725694444444\\
    0	0.153904513888889\\
    2.5	0.107626736111111\\
    5	0.0776302083333333\\
    7.5	0.0500729166666667\\
    10	0.0383125\\
    12.5	0.0431440972222222\\
    15	0.0427847222222222\\
};
\addlegendentry{LECG}

\addplot [color=\LLRmOnecolor, line width=0.8pt, mark=diamond,mark size = 1.5pt, mark options={solid, \LLRmOnecolor}]
  table[row sep=crcr]{%
    -12.5	0.375453125\\
    -10	0.339489583333333\\
    -7.5	0.290862847222222\\
    -5	0.236763888888889\\
    -2.5	0.177619791666667\\
    0	0.120027777777778\\
    2.5	0.0742916666666667\\
    5	0.0467256944444444\\
    7.5	0.03059375\\
    10	0.0168663194444444\\
    12.5	0.0146909722222222\\
    15	0.0118020833333333\\
};
\addlegendentry{LLR-$T_{h_{1}}$}

\addplot [color=\LLRmTwocolor, line width=0.8pt, mark=o,mark size = 1.5pt, mark options={solid, \LLRmTwocolor}]
  table[row sep=crcr]{%
    -12.5	0.374798611111111\\
    -10	0.338456597222222\\
    -7.5	0.289838541666667\\
    -5	0.234862847222222\\
    -2.5	0.175274305555556\\
    0	0.1174375\\
    2.5	0.0711267361111111\\
    5	0.0421510416666667\\
    7.5	0.0259375\\
    10	0.0130850694444444\\
    12.5	0.0102743055555556\\
    15	0.00730208333333333\\
};
\addlegendentry{LLR-$T_{h_{2}}$}
\addplot [color=\LLRmThreecolor, line width=0.8pt, mark=triangle,mark size = 1.5pt, mark options={solid, rotate=270, \LLRmThreecolor}]
  table[row sep=crcr]{%
    -12.5	0.374595486111111\\
    -10	0.338263888888889\\
    -7.5	0.289552083333333\\
    -5	0.234680555555556\\
    -2.5	0.174659722222222\\
    0	0.116758680555556\\
    2.5	0.0699201388888889\\
    5	0.0406545138888889\\
    7.5	0.0238385416666667\\
    10	0.0116631944444444\\
    12.5	0.00842534722222222\\
    15	0.00562847222222222\\
};
\addlegendentry{LLR-$T_{h_{3}}$}

\addplot [color=\Allcolor, line width=0.5pt, mark=10-pointed star,mark size = 1.5pt, mark options={solid, \Allcolor}]
  table[row sep=crcr]{%
    -12.5	0.372496527777778\\
    -10	0.335092013888889\\
    -7.5	0.2853125\\
    -5	0.229079861111111\\
    -2.5	0.166043402777778\\
    0	0.105128472222222\\
    2.5	0.0570208333333333\\
    5	0.0265642361111111\\
    7.5	0.0102135416666667\\
    10	0.00178472222222222\\
    12.5	0.000293402777777778\\
    15	5.20833333333333e-06\\
};
\addlegendentry{All APs}

\end{axis}
\end{tikzpicture}%

%% file: fig/FL_K_L9e25.tex
\begin{tikzpicture}

\begin{axis}[%
width=3cm,
height=4cm,
scale only axis,
xmin=10,
xmax=100,
xlabel style={font=\color{white!15!black}},
xlabel={Numb. of UEs ($K$)},
ymode=log,
ymin=1000,
ymax=5000000,
yminorticks=true,
ylabel style={font=\color{white!15!black}},
ylabel={Fronthaul load},
axis background/.style={fill=white},
title style={font=\bfseries, align=center},
xmajorgrids,
ymajorgrids,
yminorgrids,
legend style={at={(0,0.01)}, font=\footnotesize, anchor=south west, legend cell align=left, align=left, draw=white!15!black}
]

\addplot [color=\Randomcolor, line width=0.8pt, mark=x,mark size = 1.8pt, mark options={solid, \Randomcolor}]
  table[row sep=crcr]{%
10	290.8969445\\
20	1095.586778\\
30	2414.0695005\\
40	4246.345112\\
50	6592.4136125\\
60	9452.275002\\
70	12825.9292805\\
80	16713.376448\\
90	21114.6165045\\
100	26029.64945\\
110	31458.4752845\\
120	37401.094008\\
130	43857.5056205\\
140	50827.710122\\
150	58311.7075125\\
160	66309.497792\\
170	74821.0809605\\
180	83846.457018\\
190	93385.6259645\\
200	103438.5878\\
210	114005.3425245\\
220	125085.890138\\
230	136680.2306405\\
240	148788.364032\\
250	161410.2903125\\
};

\addplot [color=\LECGcolor, line width=0.8pt, mark=square,mark size = 1.5pt, mark options={solid, \LECGcolor}]
  table[row sep=crcr]{%
10	317.65625\\
20	1199.375\\
30	2645.15625\\
40	4655\\
50	7228.90625\\
60	10366.875\\
70	14068.90625\\
80	18335\\
90	23165.15625\\
100	28559.375\\
110	34517.65625\\
120	41040\\
130	48126.40625\\
140	55776.875\\
150	63991.40625\\
160	72770\\
170	82112.65625\\
180	92019.375\\
190	102490.15625\\
200	113525\\
210	125123.90625\\
220	137286.875\\
230	150013.90625\\
240	163305\\
250	177160.15625\\
};

\addplot [color=\LLSFcolor, line width=0.8pt, mark=star,mark size = 1.8pt, mark options={solid, \LLSFcolor}]
  table[row sep=crcr]{%
10	288.6778125\\
20	1086.98625\\
30	2394.9253125\\
40	4212.495\\
50	6539.6953125\\
60	9376.52625\\
70	12722.9878125\\
80	16579.08\\
90	20944.8028125\\
100	25820.15625\\
110	31205.1403125\\
120	37099.755\\
130	43504.0003125\\
140	50417.87625\\
150	57841.3828125\\
160	65774.52\\
170	74217.2878125\\
180	83169.68625\\
190	92631.7153125\\
200	102603.375\\
210	113084.6653125\\
220	124075.58625\\
230	135576.1378125\\
240	147586.32\\
250	160106.1328125\\
};

\addplot [color=\LLRmOnecolor, line width=0.8pt, mark=diamond,mark size = 1.5pt, mark options={solid, \LLRmOnecolor}]
  table[row sep=crcr]{%
10	377\\
20	1430\\
30	3159\\
40	5564\\
50	8645\\
60	12402\\
70	16835\\
80	21944\\
90	27729\\
100	34190\\
110	41327\\
120	49140\\
130	57629\\
140	66794\\
150	76635\\
160	87152\\
170	98345\\
180	110214\\
190	122759\\
200	135980\\
210	149877\\
220	164450\\
230	179699\\
240	195624\\
250	212225\\
};

%

\addplot [color=\Allcolor, line width=0.5pt, mark=10-pointed star,mark size = 1.5pt, mark options={solid, \Allcolor}]
  table[row sep=crcr]{%
10	4185\\
20	16470\\
30	36855\\
40	65340\\
50	101925\\
60	146610\\
70	199395\\
80	260280\\
90	329265\\
100	406350\\
110	491535\\
120	584820\\
130	686205\\
140	795690\\
150	913275\\
160	1038960\\
170	1172745\\
180	1314630\\
190	1464615\\
200	1622700\\
210	1788885\\
220	1963170\\
230	2145555\\
240	2336040\\
250	2534625\\
};
\addplot [color=\Randomcolor, line width=0.8pt,dotted,  mark=x,mark size = 1.8pt, mark options={solid, \Randomcolor}]
  table[row sep=crcr]{%
10	2196.72\\
20	8592.48\\
30	19187.28\\
40	33981.12\\
50	52974\\
60	76165.92\\
70	103556.88\\
80	135146.88\\
90	170935.92\\
100	210924\\
110	255111.12\\
120	303497.28\\
130	356082.48\\
140	412866.72\\
150	473850\\
160	539032.32\\
170	608413.68\\
180	681994.08\\
190	759773.52\\
200	841752\\
210	927929.52\\
220	1018306.08\\
230	1112881.68\\
240	1211656.32\\
250	1314630\\
};

\addplot [color=\LECGcolor, line width=0.8pt, dotted, mark=square,mark size = 1.5pt, mark options={solid, \LECGcolor}]
  table[row sep=crcr]{%
10	860\\
20	3320\\
30	7380\\
40	13040\\
50	20300\\
60	29160\\
70	39620\\
80	51680\\
90	65340\\
100	80600\\
110	97460\\
120	115920\\
130	135980\\
140	157640\\
150	180900\\
160	205760\\
170	232220\\
180	260280\\
190	289940\\
200	321200\\
210	354060\\
220	388520\\
230	424580\\
240	462240\\
250	501500\\  
};

\addplot [color=\LLSFcolor, line width=0.8pt, dotted, mark=star,mark size = 1.8pt, mark options={solid, \LLSFcolor}]
  table[row sep=crcr]{%
10	764.8448\\
20	2946.4592\\
30	6544.8432\\
40	11559.9968\\
50	17991.92\\
60	25840.6128\\
70	35106.0752\\
80	45788.3072\\
90	57887.3088\\
100	71403.08\\
110	86335.6208\\
120	102684.9312\\
130	120451.0112\\
140	139633.8608\\
150	160233.48\\
160	182249.8688\\
170	205683.0272\\
180	230532.9552\\
190	256799.6528\\
200	284483.12\\
210	313583.3568\\
220	344100.3632\\
230	376034.1392\\
240	409384.6848\\
250	444152\\
};

\addplot [color=\LLRmOnecolor, line width=0.8pt, dotted, mark=diamond,mark size = 1.5pt, mark options={solid, \LLRmOnecolor}]
  table[row sep=crcr]{%
10	1430\\
20	5564\\
30	12402\\
40	21944\\
50	34190\\
60	49140\\
70	66794\\
80	87152\\
90	110214\\
100	135980\\
110	164450\\
120	195624\\
130	229502\\
140	266084\\
150	305370\\
160	347360\\
170	392054\\
180	439452\\
190	489554\\
200	542360\\
210	597870\\
220	656084\\
230	717002\\
240	780624\\
250	846950\\
};

%

\addplot [color=\Allcolor, line width=0.5pt, dotted, mark=10-pointed star,mark size = 1.5pt, mark options={solid, \Allcolor}]
  table[row sep=crcr]{%
10	31625\\
20	125750\\
30	282375\\
40	501500\\
50	783125\\
60	1127250\\
70	1533875\\
80	2003000\\
90	2534625\\
100	3128750\\
110	3785375\\
120	4504500\\
130	5286125\\
140	6130250\\
150	7036875\\
160	8006000\\
170	9037625\\
180	10131750\\
190	11288375\\
200	12507500\\
210	13789125\\
220	15133250\\
230	16539875\\
240	18009000\\
250	19540625\\
};
\end{axis}

 \node [draw,fill=white] at (rel axis cs: 0.27,0.095) {\shortstack[l]{ 
    . .  {\scriptsize $L = 25$} \\
    ---  {\scriptsize $L = 9$} 
}};

\end{tikzpicture}%

%% file: fig/Numb_Net.tex
\begin{tikzpicture}
\begin{axis}[%
width=3cm,
height=4cm,
scale only axis,
xmin=5,
xmax=100,
xlabel style={font=\color{white!15!black}},
xlabel={Numb. of APs ($L$)},
ymode=linear,
ymin=0,
ymax=100,
yminorticks=true,
ylabel style={font=\color{white!15!black}},
ylabel={Number of selected APs},
axis background/.style={fill=white},
title style={font=\bfseries, align=center},
xmajorgrids,
ymajorgrids,
yminorgrids,
legend style={at={(0.01,0.37)}, font=\scriptsize, anchor=south west, legend cell align=left, align=left, draw=white!15!black}
]

\addplot [color=\Randomcolor, line width=0.8pt, mark=x,mark size = 1.8pt, mark options={solid, \Randomcolor}]
  table[row sep=crcr]{%
    9       2.2667\\
    16	    4.5750\\
    25      6.4800\\
    36	    9.6444\\
    49	    12.5878 \\
    64	    14.1499\\
    81      16.5268\\
    100	    18.9037\\
    144     21.2806\\    
};
\addlegendentry{Random}

\addplot [color=\LLSFcolor, line width=0.8pt, mark=star,mark size = 1.8pt, mark options={solid, \LLSFcolor}]
  table[row sep=crcr]{%
    9       2.2575\\
    16	    3.5075\\
    25      3.7640\\
    36	    3.8939\\
    49	    4.2451\\
    64	    5.0738\\
    81      5.4509\\
    100     5.9987\\
    144     6.4254\\
};
\addlegendentry{LLSF}

\addplot [color=\LECGcolor, line width=0.8pt, mark=square,mark size = 1.5pt, mark options={solid, \LECGcolor}]
  table[row sep=crcr]{%
    9       2.3750\\
    16	    4.0000\\
    25      4.0000\\
    36	    4.0278\\
    49	    4.6471\\
    64	    5.1637\\
    81      5.6803\\
    100     6.1968\\
    144     6.7134\\
};
\addlegendentry{LECG}

\addplot [color=\LLRmOnecolor, line width=0.8pt, mark=diamond,mark size = 1.5pt, mark options={solid, \LLRmOnecolor}]
  table[row sep=crcr]{%
    9       2.6000\\
    16	    3.7250\\
    25      5.2000\\
    36	    6.9889\\
    49	    11.0571\\
    64	    14.3014\\
    81      16.6969\\
    100     19.0925\\
    144     20.0881\\
};
\addlegendentry{LLR-$T_{h_1}$}

\addplot [color=\LLRmTwocolor, line width=0.8pt, mark=o,mark size = 1.5pt, mark options={solid, \LLRmTwocolor}]
  table[row sep=crcr]{%
    9       4.3778\\
    16	    6.5625\\
    25      10.112\\
    36	    13.144\\
    49	    22.1878\\
    64	    28.3014\\
    81      33.3938\\
    100     38.0500\\
    144     41.762\\
};
\addlegendentry{LLR-$T_{h_2}$}

\addplot [color=\LLRmThreecolor, line width=0.8pt, mark=triangle,mark size = 1.5pt, mark options={solid, rotate=270, \LLRmThreecolor}]
  table[row sep=crcr]{%
    9       4.6889\\    
    16	    7.0375\\
    25      11.032\\
    36	    14.094\\
    49	    23.7469\\
    64	    30.3014\\
    81      34.0938\\
    100     39.1500\\
    144     42.362\\
};
\addlegendentry{LLR-$T_{h_3}$}

\addplot [color=\Allcolor, line width=0.5pt, mark=10-pointed star,mark size = 1.5pt, mark options={solid, \Allcolor}]
  table[row sep=crcr]{%
    9       9\\
    16	    16\\
    25      25\\
    36	    36\\
    49	    49\\
    64	    64\\
    81      81\\
    100     100\\
    144     144\\
};
\addlegendentry{All APs}

\end{axis}
\end{tikzpicture}%

%% file: gfllr_arxiv.bbl
\begin{thebibliography}{10}

\bibitem{DiRennaAccess2020}
R.~B. Di~Renna, C.~Bockelmann, R.~C. de~Lamare, and A.~Dekorsy,
\newblock ``{Detection Techniques for Massive Machine-Type Communications: Challenges and Solutions},''
\newblock {\em IEEE Access}, vol. 8, pp. 180928--180954, 2020.

\bibitem{IAkyildizAccess2020}
I.~F. Akyildiz et~al.,
\newblock ``{6G and Beyond: The Future of Wireless Communications Systems},''
\newblock {\em IEEE Access}, vol. 8, pp. 133995--134030, 2020.

\bibitem{mmimo}
Rodrigo~C. de~Lamare,
\newblock ``Massive mimo systems: Signal processing challenges and future trends,''
\newblock {\em URSI Radio Science Bulletin}, vol. 2013, no. 347, pp. 8--20, 2013.

\bibitem{wence}
Wence Zhang, Hong Ren, Cunhua Pan, Ming Chen, Rodrigo~C. de~Lamare, Bo~Du, and Jianxin Dai,
\newblock ``Large-scale antenna systems with ul/dl hardware mismatch: Achievable rates analysis and calibration,''
\newblock {\em IEEE Transactions on Communications}, vol. 63, no. 4, pp. 1216--1229, 2015.

\bibitem{HNgoTWC2017}
H.~Q.~Ngo et~al.,
\newblock ``{Cell-Free Massive MIMO Versus Small Cells},''
\newblock {\em IEEE Trans. on Wireless Commun.}, vol. 16, no. 3, pp. 1834--1850, 2017.

\bibitem{EBjornsonTWC2020}
E.~Björnson and L.~Sanguinetti,
\newblock ``{Making Cell-Free Massive MIMO Competitive With MMSE Processing and Centralized Implementation},''
\newblock {\em IEEE Trans. on Wireless Commun.}, vol. 19, no. 1, pp. 77--90, 2020.

\bibitem{EBjornsonTCom2020}
E.~Björnson et~al.,
\newblock ``{Scalable Cell-Free Massive MIMO Systems},''
\newblock {\em IEEE Trans. on Commun.}, vol. 68, no. 7, pp. 4247--4261, 2020.

\bibitem{PopovskiAccess2018}
P.~Popovski et~al.,
\newblock ``{5G Wireless Network Slicing for eMBB, URLLC, and mMTC: A Communication-Theoretic View},''
\newblock {\em IEEE Access}, vol. 6, pp. 55765--55779, 2018.

\bibitem{TNguyenAccess2018}
T.~H. Nguyen et~al.,
\newblock ``{Optimal Power Control and Load Balancing for Uplink Cell-Free Multi-User Massive MIMO},''
\newblock {\em IEEE Access}, vol. 6, pp. 14462--14473, 2018.

\bibitem{rscf}
Andre~R. Flores, Rodrigo~C. de~Lamare, and Kumar~Vijay Mishra,
\newblock ``Clustered cell-free multi-user multiple-antenna systems with rate-splitting: Precoder design and power allocation,''
\newblock {\em IEEE Transactions on Communications}, vol. 71, no. 10, pp. 5920--5934, 2023.

\bibitem{tds}
Patrick Clarke and Rodrigo~C. de~Lamare,
\newblock ``Transmit diversity and relay selection algorithms for multirelay cooperative mimo systems,''
\newblock {\em IEEE Transactions on Vehicular Technology}, vol. 61, no. 3, pp. 1084--1098, 2012.

\bibitem{HNgoTGCN2018}
H.~Q. Ngo et~al.,
\newblock ``{On the Total Energy Efficiency of Cell-Free Massive MIMO},''
\newblock {\em IEEE Trans. on Green Commun. and Networking}, vol. 2, no. 1, pp. 25--39, 2018.

\bibitem{HDaoAccess2020}
H.~T. Dao and S.~Kim,
\newblock ``{Effective Channel Gain-Based Access Point Selection in Cell-Free Massive MIMO Systems},''
\newblock {\em IEEE Access}, vol. 8, pp. 108127--108132, 2020.

\bibitem{CDAndreaLComm2021}
C.~D’Andrea and E.~G. Larsson,
\newblock ``{Improving Cell-Free Massive MIMO by Local Per-Bit Soft Detection},''
\newblock {\em IEEE Commun. Lett.}, vol. 25, no. 7, pp. 2400--2404, 2021.

\bibitem{VRanasingheGlobecom2021}
V.~Ranasinghe, N.~Rajatheva, and M.~Latva-aho,
\newblock ``{Graph Neural Network Based Access Point Selection for Cell-Free Massive MIMO Systems},''
\newblock in {\em 2021 IEEE Global Commun. Conf. (GLOBECOM)}, 2021, pp. 01--06.

\bibitem{TVuICC2020}
T.~X. Vu et~al.,
\newblock ``{Joint Power Allocation and Access Point Selection for Cell-free Massive MIMO},''
\newblock in {\em ICC 2020 - 2020 IEEE International Conference on Communications (ICC)}, 2020, pp. 1--6.

\bibitem{TVanChienTWC2020}
T.~Van~Chien et~al.,
\newblock ``{Joint Power Allocation and Load Balancing Optimization for Energy-Efficient Cell-Free Massive MIMO Networks},''
\newblock {\em IEEE Trans. Wireless Commun.}, vol. 19, no. 10, pp. 6798--6812, 2020.

\bibitem{GDongTVT2019}
G.~Dong,
\newblock ``{Energy-Efficiency-Oriented Joint User Association and Power Allocation in Distributed Massive MIMO Systems},''
\newblock {\em IEEE Trans. Veh. Technol.}, vol. 68, no. 6, pp. 5794--5808, 2019.

\bibitem{RWangAccess2021}
R.~Wang et~al.,
\newblock ``{Performance of Cell-Free Massive MIMO With Joint User Clustering and Access Point Selection},''
\newblock {\em IEEE Access}, vol. 9, pp. 40860--40870, 2021.

\bibitem{VPalharesTVT2021}
V.~M.~T. Palhares, A.~R. Flores, and R.~C. de~Lamare,
\newblock ``{Robust MMSE Precoding and Power Allocation for Cell-Free Massive MIMO Systems},''
\newblock {\em IEEE Trans. Veh. Technol.}, vol. 70, no. 5, pp. 5115--5120, 2021.

\bibitem{UKGanesanTCom2021}
U.~K.~Ganesan et~al.,
\newblock ``{Clustering-Based Activity Detection Algorithms for Grant-Free Random Access in Cell-Free Massive MIMO},''
\newblock {\em IEEE Trans. on Commun.}, vol. 69, no. 11, pp. 7520--7530, 2021.

\bibitem{JDingCSCN2021}
J.~Ding and J.~Choi,
\newblock ``{SIC Aided $K$-Repetition for Mission-Critical MTC in Cell-Free Massive MIMO},''
\newblock in {\em 2021 IEEE Conf. on Standards for Commun. and Networking (CSCN)}, 2021, pp. 167--173.

\bibitem{AMishraCommLet2022}
A.~Mishra et~al.,
\newblock ``{Rate-Splitting assisted Massive Machine-Type Communications in Cell-Free Massive MIMO},''
\newblock {\em IEEE Commun. Lett.}, pp. 1--1, 2022.

\bibitem{ZWangWCSP2020}
Z.~Wang et~al.,
\newblock ``{NOMA in Cell-Free mMIMO Systems with AP Selection},''
\newblock in {\em 2020 Int. Conf. on Wireless Commun. and Signal Proc. (WCSP)}, 2020, pp. 430--435.

\bibitem{DiRennaWCL2019}
R.~B. Di~Renna and R.~C. de~Lamare,
\newblock ``{Adaptive Activity-Aware Iterative Detection for Massive Machine-Type Communications},''
\newblock {\em IEEE Wireless Commun. Lett.}, vol. 8, no. 6, pp. 1631--1634, 2019.

\bibitem{AGoldsmithJSAC2003}
A.~Goldsmith et~al.,
\newblock ``Capacity limits of mimo channels,''
\newblock {\em IEEE J. Sel. Areas Commun.}, vol. 21, no. 5, pp. 684--702, 2003.

\bibitem{jidf}
R.~C. de~Lamare and R.~Sampaio-Neto,
\newblock ``Adaptive reduced-rank processing based on joint and iterative interpolation, decimation, and filtering,''
\newblock {\em IEEE Transactions on Signal Processing}, vol. 57, no. 7, pp. 2503--2514, 2009.

\bibitem{spa}
R.~C. de~Lamare and R.~Sampaio-Neto,
\newblock ``Minimum mean-squared error iterative successive parallel arbitrated decision feedback detectors for ds-cdma systems,''
\newblock {\em IEEE Transactions on Communications}, vol. 56, no. 5, pp. 778--789, 2008.

\bibitem{mfsic}
Peng Li, Rodrigo~C. de~Lamare, and Rui Fa,
\newblock ``Multiple feedback successive interference cancellation detection for multiuser mimo systems,''
\newblock {\em IEEE Transactions on Wireless Communications}, vol. 10, no. 8, pp. 2434--2439, 2011.

\bibitem{dfcc}
P.~Li and R.~C. de~Lamare,
\newblock ``{Adaptive Decision-Feedback Detection With Constellation Constraints for MIMO Systems},''
\newblock {\em IEEE Trans. Veh. Technol.}, vol. 61, no. 2, pp. 853--859, 2012.

\bibitem{did}
Peng Li and Rodrigo~C. de~Lamare,
\newblock ``Distributed iterative detection with reduced message passing for networked mimo cellular systems,''
\newblock {\em IEEE Transactions on Vehicular Technology}, vol. 63, no. 6, pp. 2947--2954, 2014.

\bibitem{bfidd}
A.~G.~D. Uchoa, C.~T. Healy, and R.~C. de~Lamare,
\newblock ``{Iterative Detection and Decoding Algorithms for MIMO Systems in Block-Fading Channels Using LDPC Codes},''
\newblock {\em IEEE Trans. Veh. Technol.}, vol. 65, no. 4, pp. 2735--2741, 2016.

\bibitem{1bitidd}
Z.~Shao, R.~C. de~Lamare, and L.~T.~N. Landau,
\newblock ``{Iterative Detection and Decoding for Large-Scale Multiple-Antenna Systems With 1-Bit ADCs},''
\newblock {\em IEEE Wireless Commun. Lett.}, vol. 7, no. 3, pp. 476--479, 2018.

\bibitem{aaidd}
Roberto~B. Di~Renna and Rodrigo~C. de~Lamare,
\newblock ``Adaptive activity-aware iterative detection for massive machine-type communications,''
\newblock {\em IEEE Wireless Communications Letters}, vol. 8, no. 6, pp. 1631--1634, 2019.

\bibitem{dynovs}
Zhichao Shao, Lukas T.~N. Landau, and Rodrigo~C. de~Lamare,
\newblock ``Dynamic oversampling for 1-bit adcs in large-scale multiple-antenna systems,''
\newblock {\em IEEE Transactions on Communications}, vol. 69, no. 5, pp. 3423--3435, 2021.

\bibitem{detmtc}
Roberto~B. Di~Renna, Carsten Bockelmann, Rodrigo~C. de~Lamare, and Armin Dekorsy,
\newblock ``Detection techniques for massive machine-type communications: Challenges and solutions,''
\newblock {\em IEEE Access}, vol. 8, pp. 180928--180954, 2020.

\bibitem{comp}
Ana Beatriz L.~B. Fernandes, Zhichao Shao, Lukas T.~N. Landau, and Rodrigo~C. de~Lamare,
\newblock ``Multiuser-mimo systems using comparator network-aided receivers with 1-bit quantization,''
\newblock {\em IEEE Transactions on Communications}, vol. 71, no. 2, pp. 908--922, 2023.

\bibitem{msgamp}
Roberto B.~Di Renna and Rodrigo~C. de~Lamare,
\newblock ``Dynamic message scheduling based on activity-aware residual belief propagation for asynchronous mmtc,''
\newblock {\em IEEE Wireless Communications Letters}, vol. 10, no. 6, pp. 1290--1294, 2021.

\bibitem{msgamp2}
Roberto~B. Di~Renna and Rodrigo~C. de~Lamare,
\newblock ``Joint channel estimation, activity detection and data decoding based on dynamic message-scheduling strategies for mmtc,''
\newblock {\em IEEE Transactions on Communications}, vol. 70, no. 4, pp. 2464--2479, 2022.

\bibitem{Bradley98}
P.~S. Bradley,
\newblock ``{Feature selection via concave minimization and support vector machines},''
\newblock in {\em Proc. 13th ICML}, 1998, pp. 82--90.

\bibitem{JWangSPL2009}
J.~Wang, O.~Y.~Wen, and S.~Li,
\newblock ``{Soft-Output MMSE MIMO Detector Under ML Channel Estimation and Channel Correlation},''
\newblock {\em IEEE Signal Process. Lett.}, vol. 16, no. 8, pp. 667--670, 2009.

\bibitem{SMKay1993}
S.~M. Kay,
\newblock {\em {Fundaments of Statistical Signal Processing: Estimation Theory}},
\newblock Prentice Hall, Englewood Cliffs, NJ, 1st edition, 1993.

\bibitem{DiRennaTCom2020}
R.~B. Di~Renna and R.~C. de~Lamare,
\newblock ``{Iterative List Detection and Decoding for Massive Machine-Type Communications},''
\newblock {\em IEEE Trans. Commun.}, vol. 68, no. 10, pp. 6276--6288, 2020.

\bibitem{XWang1999}
Xiaodong Wang and H.~V. Poor,
\newblock ``{Iterative (turbo) soft interference cancellation and decoding for coded CDMA},''
\newblock {\em IEEE Trans. Commun.}, vol. 47, no. 7, pp. 1046--1061, 1999.

\bibitem{richardson}
T.J. Richardson, M.A. Shokrollahi, and R.L. Urbanke,
\newblock ``Design of capacity-approaching irregular low-density parity-check codes,''
\newblock {\em IEEE Transactions on Information Theory}, vol. 47, no. 2, pp. 619--637, 2001.

\bibitem{memd}
C.~T. Healy and R.~C. de~Lamare,
\newblock ``Design of ldpc codes based on multipath emd strategies for progressive edge growth,''
\newblock {\em IEEE Transactions on Communications}, vol. 64, no. 8, pp. 3208--3219, 2016.

\bibitem{vfap}
Jingjing Liu and Rodrigo~C. de~Lamare,
\newblock ``Low-latency reweighted belief propagation decoding for ldpc codes,''
\newblock {\em IEEE Communications Letters}, vol. 16, no. 10, pp. 1660--1663, 2012.

\bibitem{kaids}
C.~T. Healy, Zhichao Shao, R.~M. Oliveira, R.~C. de~Lamare, and L.~L. Mendes,
\newblock ``Knowledge-aided informed dynamic scheduling for ldpc decoding of short blocks,''
\newblock {\em IET Communications}, vol. 12, no. 9, pp. 1094--1101, 2018.

\bibitem{SChenJSAC2021}
S.~Chen et~al.,
\newblock ``{Structured Massive Access for Scalable Cell-Free Massive MIMO Systems},''
\newblock {\em J. Sel. Areas Commun.}, vol. 39, no. 4, pp. 1086--1100, 2021.

\bibitem{3GPPTS36814}
{\em {Further Advancements for E-ULTRA Physical Layer Aspects (Release 9)}}, document TS 36.814, 3GPP, Mar. 2017.

\end{thebibliography}
